\documentclass[conference]{IEEEtran}
\IEEEoverridecommandlockouts
\usepackage{cite}
\usepackage{amsmath,amssymb,amsfonts}
\usepackage{algorithmic}
\usepackage{graphicx}
\usepackage{textcomp}
\usepackage{epstopdf}
\usepackage{xcolor}
\graphicspath{{./Figures/}}
\DeclareGraphicsExtensions{.pdf,.jpeg,.png,.eps}
\def\BibTeX{{\rm B\kern-.05em{\sc i\kern-.025em b}\kern-.08em
    T\kern-.1667em\lower.7ex\hbox{E}\kern-.125emX}}
\begin{document}

\title{Downlink Power Control based UE-Sided Initial Access for Tactical 5G NR}

\author{\IEEEauthorblockN{Akshay Jain, Karthik Upadhya, Mikko A. Uusitalo}
\IEEEauthorblockA{\textit{Radio Systems Research} \\
\textit{Nokia Bell Labs, Espoo, Finland}\\
\{akshay.2.jain,karthik.upadhya,mikko.uusitalo\}@nokia-bell-labs.com}
\and
\IEEEauthorblockN{Harish Viswanathan}
\IEEEauthorblockA{\textit{Radio Systems Research} \\
\textit{Nokia Bell Labs, Murray Hill, NJ, USA}\\
harish.viswanathan@nokia-bell-labs.com}
}

\maketitle

\begin{abstract}
Communication technologies play a crucial role in battlefields. They are an inalienable part of any tactical response, whether at the battlefront or inland. Such scenarios require that the communication technologies be versatile, scalable, cost-effective, and stealthy. While multiple studies and past products have tried to address these requirements, none of them have been able to solve all the four challenges simultaneously. Hence, in this paper, we propose a tactical solution that is based on the versatile, scalable, and cost effective 5G NR system. Our focus is on the initial-access phase which is subject to a high probability of detection by an eavesdropper. To address this issue, we propose a novel approach that involves some modifications to the initial access procedure that lowers the probability of detection while not affecting standards compliance and not requiring any modifications to the user equipment chipset implementation. Further, we demonstrate that with a simple downlink power control algorithm, we reduce the probability of detection at an eavesdropper. The result is a 5G NR based initial-access method that improves stealthiness when compared with a vanilla 5G NR implementation.
\end{abstract}


\section{Introduction} \label{intro}

5G NR is a cost-effective communication standard owing to the economies of scale from its use in consumer networks. It is also versatile and scalable with support for multi-modal traffic, i.e., enhanced mobile broadband (eMBB), ultra-reliable low-latency communication (URLLC), and massive machine-type communication (mMTC). Tactical networks can leverage these characteristics to reduce their costs while being useful in a wide variety of tactical scenarios.

One of the key requirements placed on a communication technology in a tactical environment is the ability to maintain a low probability of detection (LPD) and intercept (LPI). However, the initial access signalling in 5G NR is designed to maximize coverage. Specifically, the initial access phase in 5G NR starts with downlink (DL) synchronization which involves the base station (gNB) transmitting a beamformed synchronization signal block (SSB) \cite{giordani2016comparative, lin20195g}. The SSB is designed to ensure coverage and be easily detectable by a UE that is trying to access the system \cite{chakrapani2020design,lin2018ss}. However, this approach is unsuitable for a tactical network since it increases the probability with which an adversary can detect and/or intercept the network. 

Several papers have offered solutions to the aforementioned issues. \cite{elmasry2021hiding} addresses the limitation of frequency hopping by randomizing the power spectral density distribution among the multiple-input multiple-output (MIMO) antenna elements to prevent the eavesdropper from determining the location/presence of a tactical network easily. Specifically, a power distribution method based on Lagrange multipliers is developed to redistribute power on all channel paths between the gNB and user equipment (UE) while reducing overall transmit power among the various multiple paths. This reduces the probability with which an eavesdropper can determine the location of the transmitter and receiver with certainty. 

Next, in \cite{cheng20225g}, physical layer techniques that are adaptations of traditional spreading, i.e., direct sequence spread spectrum (DSSS), and frequency hopping approaches, i.e., frequency hopping spread spectrum (FHSS), are considered in a tactical scenario with narrow-band internet of things (NB-IoT) (or 5G compliant low-power, low-capacity) devices. Coarse frequency hopping is accomplished in NB-IoT by scheduling the transmission resources at PRB level granularity. For DSSS, the spreading is done in the frequency domain before the inverse fast Fourier transform (IFFT) to remain compliant with the 5G orthogonal frequency division multiple access (OFDMA) waveform. This DSSS implementation will require substantial changes to the transmitter and receiver firmware. Both these techniques are shown to reduce the probability of detection (PD) or intercept (PI) and improve the anti-jamming capabilities in a 5G link, while DSSS outperforms FSSS in most cases.

A UE-sided initial-access procedure is proposed in \cite{monti2021user} and considers an eavesdropper (Eve) listening on the initial access between the gNB (Bob) and UE (Alice) to detect the presence of a tactical communication network. It is assumed that Eve can passively listen across the entire spectrum and has knowledge of the 5G protocol and associated signaling. To combat such an eavesdropper and the above mentioned challenges in initial access, \cite{monti2021user} proposes a modified 5G NR initial access where the initial-access procedure is initiated by the UE instead of the gNB. The UE transmits an omnidirectional transmission that contains an initial sequence protected by a \textit{nonce} to inform its presence to the gNB. After the gNB detects this initial transmission, it estimates the direction of arrival (DoA) of the transmitted signal, and based on the estimated DoA, it determines a reduced set of SSB beams for the initial access procedure. This approach ensures that the SSB beams are transmitted in the direction of the UE rather than over the entire coverage area of the cell. However, the challenge with this method is that it requires modification of the firmware of a commercial off-the-shelf (COTS) UE, which is generally difficult to achieve due to the maturity of existing products as well as high inertia from chip manufacturers. Another limitation of the analysis in \cite{monti2021user} is that the simulations are limited to a simple additive white Gaussian channel without frequency selectivity or interference. There is also the limitation that a purely direction-aided approach requires a large antenna array at the gNB to be very effective, and such requirements cannot typically be met with portable solutions at the tactical edge. Additionally, it assumes that the gNB and UE are perfectly time-synchronized.  

Hence, in this article we propose:
\begin{itemize}
	\item A novel UE-sided initial-access system and method, that is compliant with 3GPP standards and can be implemented using a COTS UE. Note that, while \cite{monti2021user} proposes a UE-sided initial-access method, it requires substantial modifications to the UE hardware and software to accomplish the same. Such a method, which requires deeper modifications on the UE side, is costly and requires larger time-to-market.  On the other hand, our proposed method can be used with a COTS UE without any major modifications, thus reducing the CAPEX and OPEX for the end-user as well as ensuring faster time-to-market for the vendor.
	\item A novel LPD/LPI approach that utilizes DL power-control when transmitting the DL synchronization signals. 
\end{itemize}

The article is organized as follows: The DL synchronization signaling in 5G NR initial access is introduced in Section \ref{sec:5GNRInitialAccess}. In Section \ref{sec:auxUeInitialAccess}, the proposed UE sided initial access procedure incorporating downlink power control is discussed in detail.  Results and discussions from the simulation studies conducted are presented in Section \ref{sec:simulationResults} and Section \ref{sec:conclusion} concludes the paper.

\section{5G NR Downlink Synchronization and its Challenges}
\label{sec:5GNRInitialAccess}

The initial access procedure in 5G NR comprises of two parts: downlink (DL) and uplink (UL) synchronization. The gNB initiates DL synchronization over the broadcast channel, and the UE listens to these DL synchronization signals to extract essential information from the synchronization signals (SS) and the master information block (MIB). Following this, the UE initiates UL synchronization via the random-access channel (RACH) procedure.

For DL synchronization, the gNB broadcasts the synchronization signal block (SSB) which is a combination of the synchronization signals (SS) and physical broadcast channel (PBCH). The SS consists of two synchronization signals, i.e., the primary SS (PSS) and secondary SS (SSS) which are binary phase-shift keying (BPSK) modulated m- and Gold-sequences, respectively, with a sequence length of 127 \cite{chakrapani2020design,lin2018ss}. The PSS is generated with the parameter $ \mathrm{NID2} $ and the SSS is generated with both $ \mathrm{NID1} $ and $ \mathrm{NID2} $ \cite{chakrapani2020design,lin2018ss} where  $\mathrm{NID1}\in \left\{0,1,\ldots,335\right\}$ and  $\mathrm{NID2}\in\left\{0,1,2\right\}$ together form the physical cell ID (PCI).

\begin{figure}[htb] 
	\centering
	\includegraphics[width=0.5\columnwidth]{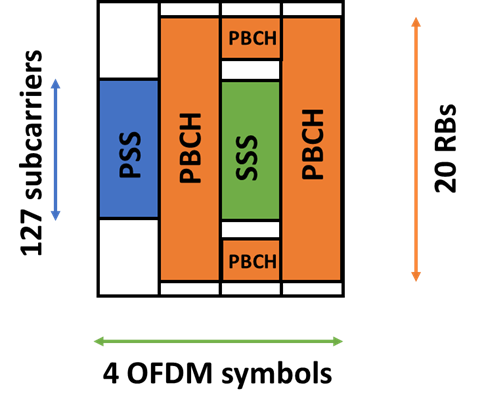}
	\caption{SS Block time-frequency grid structure }
	\label{fig:ssbTimeFrequencyGridStructure}
\end{figure}

The time-frequency grid structure of a SSB is presented in Fig. \ref{fig:ssbTimeFrequencyGridStructure}. The SSB spans $ 4 $ OFDM symbols in the time domain and $ 20 $ resource blocks (RBs) in the frequency domain. The first symbol contains only the PSS and is used by the UE for detecting the SSB, inferring $ \mathrm{NID2} $, and performing coarse frequency-estimation. The third symbol in an SSB contains the SSS surrounded by the PBCH. The UE utilizes the SSS to infer $ \mathrm{NID1} $ and evaluate the reference signal received power (RSRP). The estimated PCI is then used to decode the PBCH.

The broadcasted SSB is part of an SSB burst which contains $ P $ repetitions of the SSB with each transmission done on one of the $ P $ different beams. $ P $ can be upto $ 4 $ or $ 8 $ at sub-6 GHz frequencies, while it can be upto $ 64 $ at mmWave frequencies. The beamforming is done to utilize the array gain at the gNB to increase coverage for cell-edge UEs.

\begin{figure}[htb] 
	\centering
	\includegraphics[width=\columnwidth]{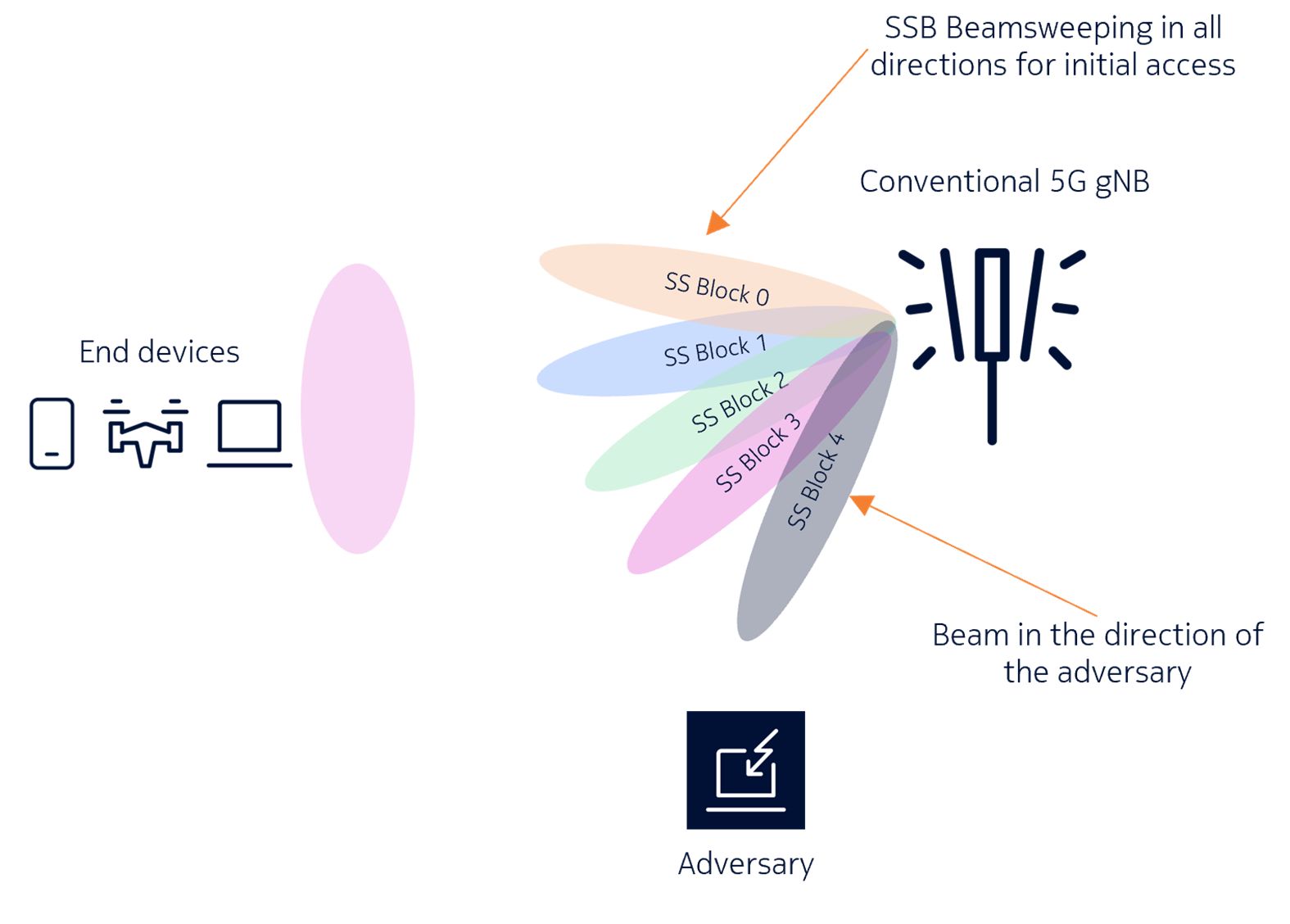}
	\caption{Baseline 5G NR system}
	\label{fig:baseline5GnrSystem}
\end{figure}

Notably, the baseline 5G NR scenario involves performing beam-sweeping/scanning across the entire cell-sector to increase coverage, but which can inadvertently result in a beam pointing towards the eavesdropper. The eavesdropper can then utilize the publicly available knowledge about the structure of SSB signals, i.e., the PSS and SSS, to detect the presence of a tactical network. This scenario is illustrated in Fig. \ref{fig:baseline5GnrSystem}.

\begin{figure}[htb] 
	\centering
	\includegraphics[width=\columnwidth]{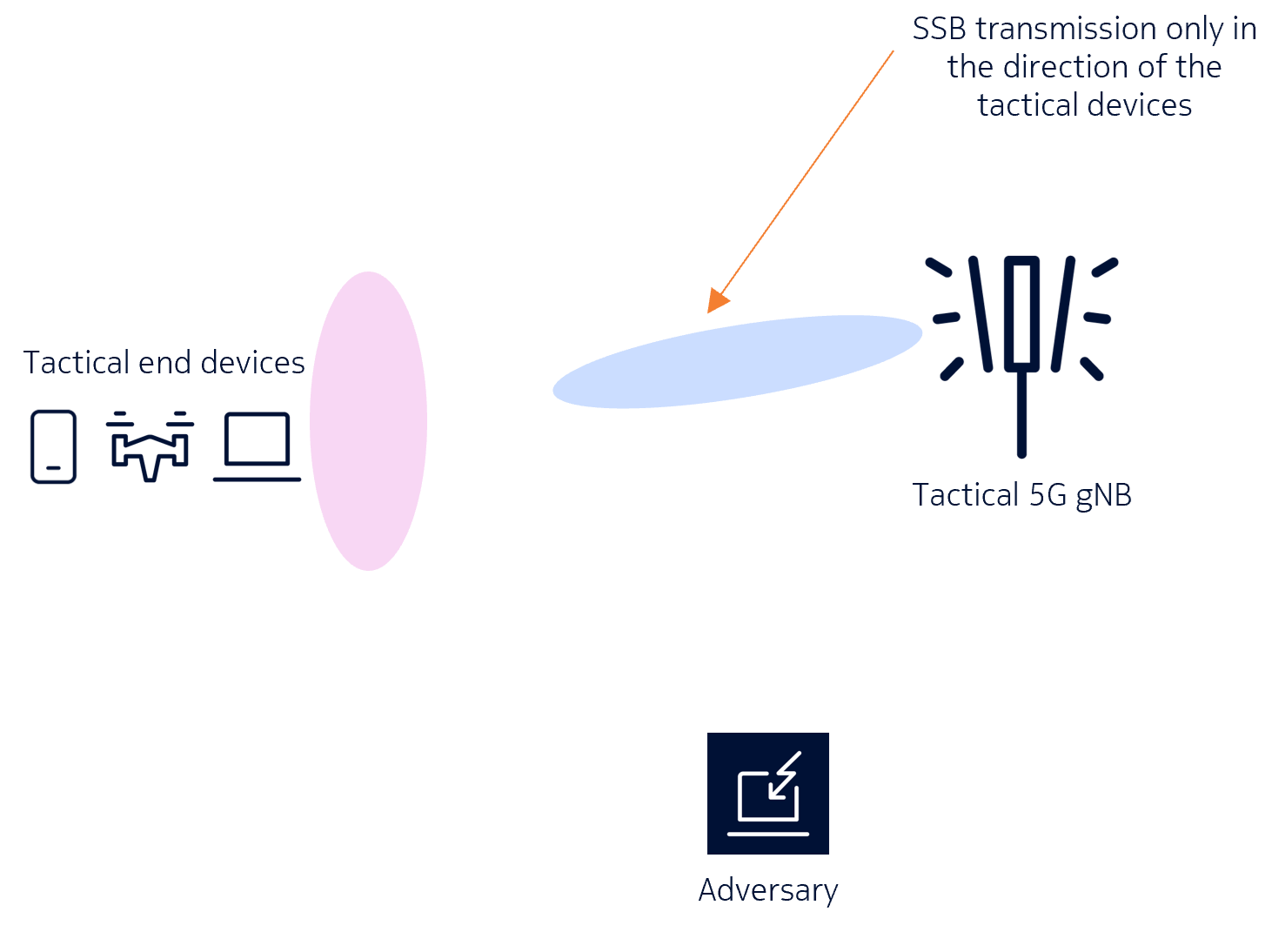}
	\caption{Desired 5G NR system behavior, where the baseline system sends the SSB beam only in the direction of the UE}
	\label{fig:tacticalNetwork}
\end{figure}

On the other hand, the behavior that a tactical network should ideally exhibit is presented in Fig. \ref{fig:tacticalNetwork}. Specifically, the tactical gNB should transmit the SSB towards only the legitimate tactical UE thereby reducing the likelihood that the signal is detected by an eavesdropper. This requirement necessitates that the gNB DL synchronization procedure be modified for lower PD and PI while ensuring that the resulting system works seamlessly with a COTS UE. We solve this problem with an auxiliary-UE based system for initial access that is detailed in the Section \ref{sec:auxUeInitialAccess}.

\section{Auxiliary UE based Initial Access} 
\label{sec:auxUeInitialAccess}

In order to reduce the probability of detection and intercept of 5G NR initial access without any hardware modifications of a COTS UE, we propose the addition of a collocated auxiliary UE (aux-UE) to the COTS UE. The aux-UE can be a software defined radio (SDR) or any other standardized or proprietary implementation of a transmitter and receiver. It is connected to the COTS UE through a commercially standardized wired or wireless link such as USB Type-C, Bluetooth, or wi-fi. 

\begin{figure} 
\centering
\includegraphics[width=\columnwidth]{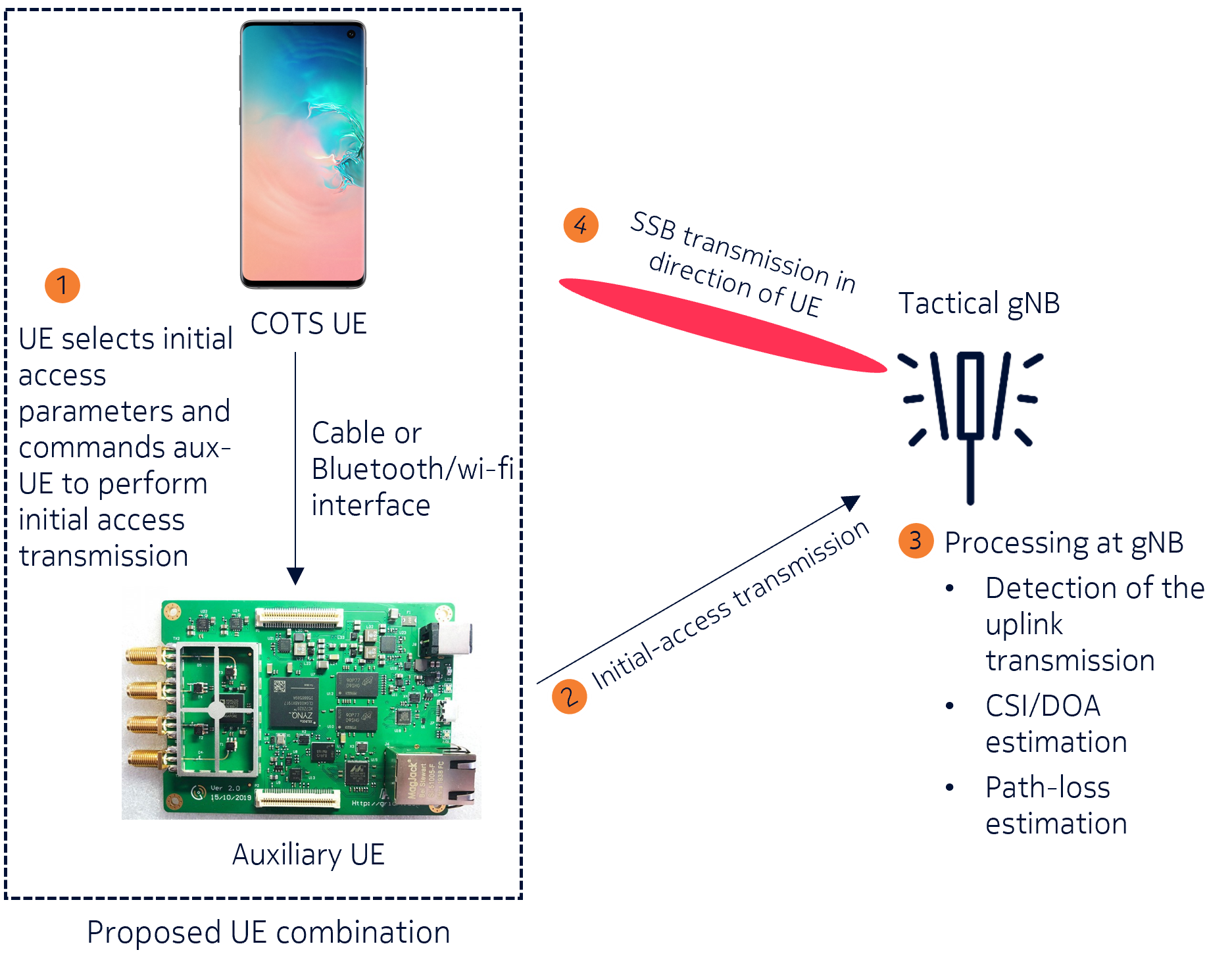}
\caption{Initial access procedure with an aux-UE}
\label{fig:proposedSystem}
\end{figure}

With this setup, the initial access procedure is initiated by the UE instead of the gNB, as illustrated in Fig. \ref{fig:proposedSystem}. During startup/initial access, the UE selects some initial-access parameters such as the Cell-ID to be used by the gNB\footnote{By selecting the cell-ID to be used by the gNB, the tactical UE can perform cell-ID hopping without a priori exchanging the hopping pattern with the gNB.}, the aux-UE transmit power, and a \textit{nonce} for authentication. The UE instructs the aux-UE to send an initial access transmission with a custom sequence including these parameters. The tactical gNB then performs the following tasks:
\begin{itemize}
	\item \textbf{Detection:}The gNB detects the transmission by correlating the received waveform with the known uplink waveform. It also decodes the parameters transmitted by the UE.
	\item \textbf{Channel state information (CSI) or direction of arrival (DoA) estimation:} The gNB estimates the CSI between the aux-UE and the gNB. Alternatively, it may also perform DoA estimation to determine the direction of the aux-UE transmission relative to the gNB. Note that, here it is assumed that the aux-UE and COTS UE are collocated. Hence, the CSI/DoA estimation of the aux-UE transmission leads to an almost accurate estimation of the correct beam for the COTS UE of interest.\footnote{Note that CSI/DoA estimation is not performed in this paper, but instead we perform genie-aided beam selection detailed in Section \ref{sec:simulationResults}.}
	\item \textbf{RSRP estimation:} The gNB also determines the RSRP, which given the aux-UE transmit power, can be used to estimate of the path-loss between the UE and the gNB.
\end{itemize}

Using the CSI/DoA estimates, the gNB then transmits the SSB only in the direction of the aux-UE instead of performing the beamsweeping procedure. The transmit power of the SSB transmission is adjusted based on the RSRP.

\subsection{Downlink Power Control} 
\label{sec:downlinkPowerControl}
An approach to reduce the PD and PI is to utilize DL power control at the gNB. By suitably adjusting the DL transmit power, we can reduce the probability with which the SNR at the eavesdropper is large enough for it to successfully detect the SSB, while at the same time, the probability of detection and decode of the SSB at the tactical UE is unaffected.
 
For this process, the gNB utilizes the information provided by the aux-UE to estimate the RSRP. Specifically, the gNB utilizes the aux-UE transmit power and the estimated RSRP to determine the path-loss and compute the SSB transmit power such that the SNR of the received SSB meets a target SNR $S_{target}$ at the tactical UE. This transmit power is computed in a straightforward manner as
\begin{equation}
P_{\mathrm{tx}} = \eta_{\mathrm{tue}} - \Gamma_{\mathrm{tue}} + S_{\mathrm{target}}
\label{eqn:rxPowerEqn}
\end{equation}
where $P_{\mathrm{tx}}$ refers to the transmit power used by the tactical gNB, $\eta_{\mathrm{tue}}$ is the noise power at the UE receiver, $\Gamma_{\mathrm{tue}}$ is the path-gain between the tactical gNB and UE. The value $S_{target}$ is selected such that the UE is guaranteed to detect and decode the SSB and PBCH.

\section{Simulation Results} 
\label{sec:simulationResults}

This section presents simulation results demonstrating the benefits of an aux-UE system with downlink power control in reducing the PD. For the analysis, we consider the 5G NR system as the baseline and compare its PD performance with that of a tactical 5G NR system that has a COTS UE linked to an aux-UE as shown in Fig. \ref{fig:proposedSystem}. Furthermore, for the proposed solution we assume that the gNB transmits the SSB on the ``best'' beam, i.e., we assume a genie-aided SSB beam-selection algorithm that picks a beam that results in the highest RSRP at the UE\footnote{In practice, the optimal beam will have to be estimated based on the direction from which the gNB receives the signal transmitted by the aux-UE.}. The considered assumption is valid because the PD, when considering aux-UE transmission and DL synchronization, is determined by the PD of DL synchronization due to the higher DL transmit power vis-a-vis the UL transmit power, known SSB signal structure, and the beamsweeping/scanning procedure. Consequently, the assumption of a genie-aided SSB beam-selection will give us a lower bound on detection probability for the eavesdropper during DL synchronization. The capabilities of both the systems are specified in Table \ref{table:capabilityDescription}. The eavesdropper is assumed to have the capabilities of a COTS UE. 
\begin{table}[htb]
\caption{Description of capabilities of the simulated baseline and the proposed tactical 5G system}
\begin{center}
\begin{tabular}{|p{4cm}|p{4cm}|}
\hline
\textbf{Baseline 5G NR system}&\textbf{Proposed Tactical 5G NR system} \\ \hline 
Conventional standard compliant gNB and UE & Conventional standard compliant UE. Aux-UE attached for initial access \\ \hline
gNB transmits SSB in all directions & UE transmits initial access request using the aux-UE. gNB responds with an SSB in the UE direction \\ \hline
SSB periodicity fixed and set to default value of 20ms & SSB periodicity fixed and set to default value of 20ms\\ \hline
All beams scanned in each SSB burst & Only the most suitable beam transmitted in the direction of the UE \\ \hline
All beams transmitted without repetition& The most suitable beam transmitted with equal power without repetition\\ \hline
Fixed cell-ID unknown to UE. UE will scan through all possible combinations of $ {\mathrm{NID1} \in \left\{0, 1, \ldots, 335\right\}} $ and $ \mathrm{NID2} \in \left\{0, 1, 2\right\} $. $\text{Cell ID} = 3\mathrm{NID1} + \mathrm{NID2}$ & Cell-ID known to vanilla 5G UE through AT commands and changing with each SSB burst\\ \hline
\end{tabular}
\label{table:capabilityDescription}
\end{center}
\end{table}

Table \ref{table:simulationParameters} presents the simulation scenario and its related parameters. For the simulations, the gNB transmit power is obtained by matching the cell-edge RSRP/SNR with that of real-world deployments. We assume a hexagonal grid of cells placed at an inter-site distance of $ 200 $m. This benchmarking is done using a system-level simulator with the 3GPP 38.901-UMi channel model and resulted in a gNB transmit power of $ 28 $ dBm for a signal bandwidth of $ 20 $ PRBs (corresponding to the SSB) and the already specified inter-site distance.  

We evaluate the PD at the eavesdropper with both an energy detector and a correlation-based receiver. For the energy detector, the eavesdropper is assumed to not have the knowledge of the SSB signal structure nor its location in time or frequency. Instead, the eavesdropper scans across the observation bandwidth and time window specified in Table~\ref{table:simulationParameters}. The scanning process is done with a sliding window of width $ 20 $ PRBs in frequency and a duration of $ 4 $ OFDM symbols in time (assuming a subcarrier spacing same as that of the tactical system) over the time-frequency observation window. The detector computes the received energy for each position of the sliding window and declares the existence of the DL synchronization signal when the energy in any one of these positions exceeds a threshold. The thresholds are varied to plot the entire receiver operating characteristic (ROC).

On the other hand, it is assumed that the eavesdropper has knowledge of the SSB signal structure for the correlation-based approach. However, similar to the energy detector, this receiver does not assume to know the location of the PSS or SSS in time or frequency and instead, scans across the observation bandwidth and time window. For each value of time and carrier frequency offset (the former is at sample spacing and the latter is at subcarrier spacing), the receiver correlates the received signal with the PSS and SSS and declares the existence of the DL synchronization signal when the output of the correlator exceeds a threshold at any combination of time or carrier frequency offset in the window. Similar to the energy detector, here too the thresholds are varied to plot the entire ROC.

\begin{figure} 
\centering
\includegraphics[width=\columnwidth]{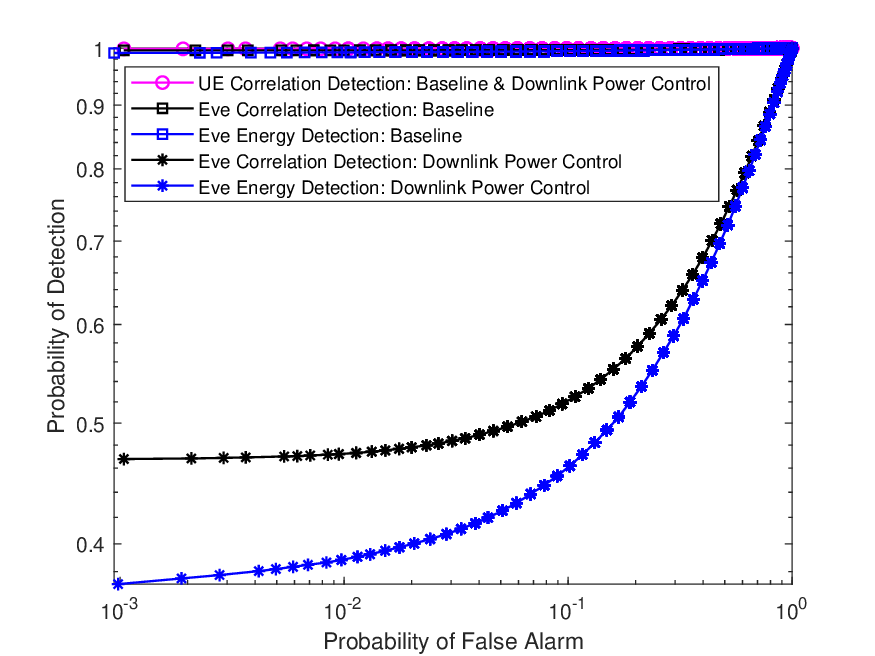}
\caption{ROC for UE and eavesdropper in the baseline scenario and proposed scenario (UE-sided initial access with genie-aided SSB beam selection and DL power control) }
\label{fig:powerControl}
\end{figure}

\begin{table}[htb]
\caption{Simulation Parameters}
\begin{center}
\begin{tabular}{|p{4.5cm}|p{3.5cm}|}
\hline
\textbf{Parameter Name}&\textbf{Value} \\ \hline 
Number of cells & $1$ \\ \hline
Number of cell sectors & $ 3 $ \\ \hline
Sector width & $ 120^\circ $  \\ \hline
Number of UEs & $ 1 $ \\ \hline
UE antenna array configuration $ {(\text{row} \times \text{column} \times \text{polarization})} $ & $ 2 \times 1 \times 2 $ \\ \hline
UE location & Randomly distributed across the cell \\ \hline
gNB antenna array configuration $ {(\text{row} \times \text{column} \times \text{polarization})} $  & $ 4 \times 2 \times 2 $ \\ \hline
gNB transmit power & $ 28 $ dBm \\ \hline
SSB transmission period & $ 20 $ ms \\ \hline
Number of Eavesdropper & $ 1 $ \\ \hline
Eavesdropper Antenna configuration $ {(\text{row} \times \text{column} \times \text{polarization})} $ & $ 2 \times 1 \times 2 $\\ \hline
Eavesdropper location & Randomly distributed across the cell\\ \hline
Eavesdropper observation bandwidth & $ 15.36 $ MHz \\ \hline
Eavesdropper observation time & $ 25 $ ms \\ \hline
Channel model & 3GPP 38.901 UMi \\ \hline
Carrier frequency & $ 3.5 $ GHz \\ \hline
Cell inter-site distance & $ 200 $ m \\ \hline
SNR target $ (S_{\mathrm{target}}) $ & $ 0 $ dB \\ \hline
\end{tabular}
\label{table:simulationParameters}
\end{center}
\end{table}
Fig. \ref{fig:powerControl} shows the detection performance, through the ROC, of a tactical UE and an eavesdropper for the baseline scenario and the proposed solution with genie-aided beam selection and downlink power control. The probabilities of detection and false alarm are averaged over $ 1000 $ random eavesdropper and UE locations, wherein at each drop the results are further averaged over $50$ different channel realizations. Additionally, in the baseline scenario all three cell sectors transmit the DL synchronization signals. On the other hand, for the proposed scenario wherein the gNB can measure the received RSRP from aux-UE's UL transmission, only the cell sector with the best received RSRP transmits the DL synchronization signals. This cell sector, as a consequence, also corresponds to the sector in which the UE is located. 

In Fig. \ref{fig:powerControl}, for the baseline scenario where there is no power control, it can be observed that the eavesdropper's detection performance approaches that of the tactical UE. This is the case with both the energy detector and correlator and it is because of the high transmit power and the fat SSB beams owing to the small number of antennas at the gNB. Note that, even though the difference is minor, the energy detector performs worse than the correlator because it does not utilize the signal structure. It can also be inferred that 3GPP based 5G NR SSB beam does not offer the LPD characteristic essential for tactical networks. 

% despite the fact that the gNB is transmitting only on a single beam in the direction of the UE
%\begin{figure} 
%\centering
%\includegraphics[width=\columnwidth]{nopwrctrl}
%\caption{ROC with genie-aided UE-side initial access without DL power control}
%\label{fig:noPowerControl}
%\end{figure}

\begin{figure} 
\includegraphics[width=\columnwidth]{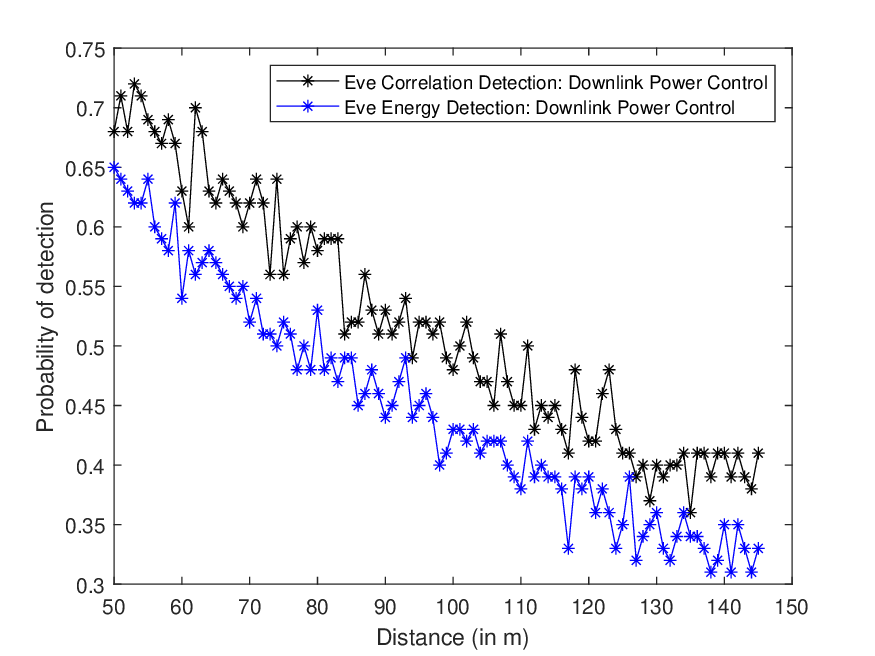}
\caption{Plot of the probability of detection for the eavesdropper at different distances from the gNB. The probability of false alarm is set to $ 10\% $.}
\label{fig:distancePlot}
\end{figure}
 
In contrast, from Fig. \ref{fig:powerControl} it can also be observed that the PD for the scenarios with DL power control, which also encompasses the correct SSB beam-selection, provisions a substantial reduction of approximately $ 50\%$-$60\% $ in detection performance for the eavesdropper. However, the UE detection performance is unaffected and is similar to the baseline scenario. In effect, by ensuring that the gNB scales down its transmit power to a level which is sufficient for the tactical UE to detect and decode the DL synchronization signals, we have changed the regime where 5G NR operates from one that ensures coverage to one that improves stealth without any substantial changes to the UE-side hardware. 

Lastly, It must be emphasized that the results presented in Fig. \ref{fig:powerControl} are for the case when the tactical UE and eavesdropper are dropped within the same cell. This is typically not the case in tactical scenarios since the eavesdropper would want to be out-of-sight from the tactical network operators. Hence, for the scenario where the eavesdropper is further away, for example at the cell edge, it can be observed from Fig. \ref{fig:distancePlot} that for a false alarm rate of $ 10\% $ the detection probability is less than $ 50\% $. As the eavesdropper moves further away, for the same false alarm rate, the detection probability gets worse. 

Thus, it can be inferred that an aux-UE-based initial-access system combined with DL power control results in a tactical 5G NR system with lower PD, whilst ensuring a low-cost, scalable, and faster time-to-deployment implementation.  

\section{Conclusion}
\label{sec:conclusion}
We consider 5G NR as a candidate for tactical networks and focus on its stealthiness. We quantify this attribute by the probability with which an eavesdropper can detect or intercept the network. Our interest is in a solution that makes 5G NR stealthy while not requiring extensive modifications to the COTS UE, thereby enabling us to preserve attributes that make 5G NR attractive for use in a tactical network, namely, standards compatibility and economies of scale. 

In this regard, a novel aux-UE based UE-side initial-access approach for a tactical 5G NR system was presented in this paper. A DL power-control approach was also introduced and was shown to reduce the detection probability by $ 50\% $-$ 60\% $ for the eavesdropper without sacrificing the performance of the tactical UE. Lastly, the impact of distance on the eavesdropper detection probability has also been presented, indicating the advantages the proposed approach has when the eavesdropper is at the cell edge or further away even in relaxed false alarm rate requirements. 

In future work, sophisticated initial-access sequence-design and improved DL SSB beam-design approaches will be investigated to further lower the PD/PI for the proposed tactical 5G NR system. 
\bibliographystyle{IEEEtran}  
\bibliography{references}  
\end{document}